\documentclass[12pt]{article}
\usepackage{amsmath}
\usepackage{amssymb,amsthm}
\usepackage{color}

\newtheorem{Thm}{Theorem}[section]
\newtheorem{theorem}[Thm]{Theorem}
\newtheorem{proposition}[Thm]{Proposition}

\title{On Schr\"{o}dinger's solution of the EPR-locality paradox for nonlocal and local measurements}

\author{Walter F. Wreszinski\footnote{wreszins@gmail.com, 
Instituto de Fisica, Universidade de S\~ao Paulo (USP), Brasil}}        
        
\begin{document}

\maketitle

\begin{abstract}

We review Schr\"{o}dinger's solution of the EPR-locality paradox \cite{Schr}, emphasizing that it consists in applying what we call
``Schr\"{o}dinger's principle'' to systems defined on Fock space, in the case of nonlocal measurements, such as values of spin components,
or of directions of photon polarizations. We suggest, on the basis of a specific local measurement, that of a local current in relativistic
quantum field theory, which features might replace Schr\"{o}dinger's principle. We also review the fact that a full answer remains open even 
in the framework of formal series in perturbation theory, due to the Lieb-Loss inequalities. 

\end{abstract}

\section{Introduction}

The present year marks the centennary of (Schr\"{o}dinger) quantum mechanics, which originates on his papers in the journal Annalen der
Physik (the first one being Ann. Phys. 79, 361 (1926)). Today, it seems of both historical and philosophical interest to inquire on his
approach to the conceptual foundations of quantum mechanics. In the words of Wigner \cite{Wi}, ``there are two basic concepts in quantum 
mechanics: states and observables. The states are vectors in Hilbert space, the observables self-adjoint operators acting on these 
vectors.'' We may ask whether any doubts remain about the general validity of this framework: for instance, is the EPR-locality paradox
\cite{EPR} cleared up?

In the present paper we attempt to complete some steps in Schr\"{o}dinger's analysis \cite{Schr} which allow a complete proof of the
non-existence of the EPR paradox in the case of measurements of nonlocal quantities, such as spin components or photon polarizations, 
within the Copenhagen interpretation. Griffiths \cite{Gri1} has arrived at the same conclusion within his own probabilistic interpretation, 
see also \cite{Gri2}. The issue of \textbf{local} measurements is an entirely different one, see section 4 for references and open problems.

The main issues occur already in the case of quantum systems with finite number
of degrees of freedom (ndof), for which there are only two guiding principles which rule the structure of wave-vectors: the building of
tensor products and the superposition principle. The principle which we suggest to call \textbf{Schr\"{o}dinger's principle} concerns the 
formation of tensor products. An obvious generalization concerns free quantum fields, defined on Fock space, which generalizes this structure
to include systems with infinite ndof.

In fact, as we shall see, the (so-called) EPR-locality paradox is well-posed already in the simplest example of states of two spins one-half. 
In general, for a finite region $\Lambda \subset \mathbf{Z}^{\nu}$, where $\nu$ denotes the dimension, the  Hilbert space is given by 
${\cal H}_{\Lambda} = \otimes_{x \in \Lambda}\mathbf{C}_{x}^{2}$, and the set of observables is the set ${\cal A}(\Lambda) = B({\cal H}_{\Lambda})$
of bounded operators on this Hilbert space, generated by the set of operators $S^{i}_{x}=1/2 \sigma^{i}_{x}$, where $i=1,2,3$ and 
$\sigma^{i}_{x}$ denote the Pauli matrices at the site $x$.

\section{Generalities on states and observables}

Let $\rho_{\Lambda}$ denote a density matrix on ${\cal H}_{\Lambda}$, that is, a positive, Hermitean matrix of unit trace. A \emph{state} 
$\omega_{\Lambda}$ on ${\cal A}(\Lambda)$ is a positive, normed linear functional on ${\cal A}(\Lambda)$:
$\omega_{\Lambda}(A) = Tr_{{\cal H}_{\Lambda}} (\rho_{\Lambda} A) \mbox{ for } A \in {\cal A}(\Lambda)$
(positive means $ \omega_{\Lambda}(A^{\dag}A) \ge 0$, normed $\omega_{\Lambda}(\mathbf{1})=1$.). 

A state $\omega_{\Lambda}$ on the observable algebra , with the assumption that $\Lambda_{1}$ and $\Lambda_{2}$ are \textbf{disjoint} regions, 
${\cal A}_{\Lambda_{1}} \cup {\cal A}_{\Lambda_{2}}$ is a \textbf{product state} iff there are states $\omega_{\Lambda_{1}}$ on ${\cal A}_{\Lambda_{1}}$
and $\omega_{\Lambda_{2}}$ on ${\cal A}_{\Lambda_{2}}$ such that
$$ 
\omega_{\Lambda}(XY) = \omega_{\Lambda_{1}}(X) \omega_{\Lambda_{2}}(Y) \forall X \subset {\Lambda_{1}} \mbox{ and } Y \subset \Lambda_{2} 
$$
A state is called \textbf{separable} iff it is a convex combination of product states. Otherwise it is called \textbf{entangled}.
The former states are thereby ``classical'' in the sense of exhibiting ``product correlations'', and the entangled states, in the 
complementary set, are supposedly the most ``intrinsically quantum''.  Notable examples of entangled states called Bell states
\cite{Li}, one of the most famous of them being
\begin{equation}
\label{(1.1)}
\omega_{B}(.) \equiv (\Psi_{B,-}, . \Psi_{B,-}) \mbox{ where } \Psi_{B-} \in \mathbf{C}_{1}^{2} \otimes \mathbf{C}_{2}^{2}
\end{equation}
where
\begin{equation}
\label{(1.2)}
\Psi_{B,-} \equiv \frac{1}{\sqrt(2)} (|+)_{1} \otimes |-)_{2}-|-)_{1} \otimes |+)_{2}) 
\end{equation}
and $|\pm)_{1,2}$ denote the basis of the two orthonormal eigenvectors of $\sigma^{3}$ on the spaces $\mathbf{C}_{1}^{2}$ resp. $\mathbf{C}_{2}^{2}$.

Quantum spin systems are similar to quantum fields because of the now famous Lieb-Robinson bound \cite{LR}: for $A,B \in {\cal A}_{0}$, where
${\cal A}_{0}$ denotes the algebra of observables at the origin, and $\tau_{x}$ and $\tau_{t}$ denote the so-called space-translation and 
time-translation automorphisms of the so-called algebra of quasi-local observables of the infinite system \cite{BRo}, which may be viewed as
(suitable) extensions of the unitary space and time-evolutions of the algebras of the finite system:
\begin{equation}
\label{(1.3)}
||[(\tau_{x}\tau_{t})(A),B]|| \le 2 C ||A||||B|| \exp[-|t|(\lambda |x|/|t|- D)]
\end{equation}
where $C, \lambda >0, D$ are quantities which depend on the interaction.
In ~\eqref{(1.3)},$(\tau_{x}\tau_{t})$ may be replaced by $(\tau_{t}\tau_{x})$. The commutator $[(\tau_{x}\tau_{t})(A),B]$ with
$B \in {\cal A}_{0}$ provides a measure of the dependence of the observation $B$ at the point $x$ at time $t$ at the origin at time $t=0$,
showing that this effect decreases exponentially with time outside the cone
\begin{equation}
\label{(1.4)}
|x| < v_{g}|t| 
\end{equation}
where the quantity $v_{g}$ is called the \textbf{group velocity}, which satisfies
\begin{equation}
\label{(1.5)}
0<v_{g}< \infty
\end{equation}
for a large class of interactions, which include the finite-range interactions (see \cite{BRo}, p. 254).

\section{The EPR-locality paradox and Schr\"{o}dinger's principle}

We refer to the observers associated to the ``soap-opera'' (\cite{Li}, p.34) as Alice, who performs measurements of $\sigma^{3}_{2}$ and Bob, 
who measures $\sigma^{3}_{1}$. In addition, Charlie, a third character in the opera, introduced by Griffiths (\cite{Gri2}, p.19), prepares 
a spin singlet at the center of the laboratory and sends the two particles in opposite directions towards two apparatuses 1 and 2, constructed, 
respectively, by Alice and Bob, \textbf{who are both aware of the initial state ~\eqref{(1.2)}}. This spin system has no dynamics, but its 
observables may be considered as part of an infinite variety of (infinite) quantum spin system, with a definite dynamics, see, e.g., \cite{Wre1}.
Take any one of them as an example, the results are all, of course, equivalent, because they depend only on the validity of the Lieb-Robinson
bound ~\eqref{(1.3)}. The corresponding  symmetries of time translation and space translation are, then, described by so-called automorphisms 
$\tau_{t}$ and $\tau_{x}$, respctively, of an algebra ${\cal A}$ of the infinite system, which are one-to-one mappings of ${\cal A}$ onto itself, 
which preserve the algebraic structure. Above, $t \in \mathbf{R}$ and $x \in \mathbf{Z}^{\nu}$, where $\nu$ denotes the dimension.  

We now ask: in what consists the EPR-locality paradox \cite{EPR}, and is it well posed? The paradox consists in the following consideration:
if Alice performs a measurement and finds a certain value of the (3-component) of the spin, she is able to deduce the value that Bob will
be measuring, even if he is beyond the reach of any signal, due to the fact that she knows the initial state (~\eqref{(1.1)} or ~\eqref{(1.2)}).

Notice that in ~\eqref{(1.2)}, the two basic building blocks of quantum mechanics of finite ndof- tensor products and the superposition 
principle- are present: we refer to Lieb's \cite{Li} stimulating lectures on the subject, which we follow in part. About this state, 
Schr\"{o}dinger \cite{Schr} writes: ``by their interaction, the two representatives (or functions) have become entangled.''. He says, 
in addition, that ``entanglement is not one, but \textbf{the} characteristic trait of quantum mechanics, the one that enforces its entire 
line of departure from classical lines of thought''.

We formulate Schr\"{o}dinger's description of this ``trait'' as an independent principle:

\textbf{Schr\"{o}dinger's principle}: The characteristic ``trait'' of quantum mechanics is that two physical systems that are brought into
temporary interaction and then completely separated, can no longer be described, \textbf{even after the interaction has utterly ceased}, by
individual wave-functions for each of the two systems.

The word ``interaction'' used by Schr\"{o}dinger in Schr\"{o}dinger's principle refers to the formation of an entangled state such as 
~\eqref{(1.1)}, the ``complete separation'' is the operation performed by Charlie. We have, now:

\begin{proposition}
\label{prop:1.1}
The EPR-locality paradox is well posed under the following assumptions:
1) the time $T$ of a quantum mechanical measurement satisfies $T<\infty$ (\cite{Wre2})
If, in addition, we assume that
2) at the moment the measurement is performed, there occurs the collapse of the wave-packet 
then the EPR-locality paradox disappears as soon as Schr\"{o}dinger's principle is invoked.

\begin{proof}

Under assumption 1), if the initial state is subject to a dynamics satisfying the assumptions of the Lieb-Robinson bound ~\eqref{(1.3)},
there is a finite group velocity FGV ~\eqref{(1.4)}, and therefore Bob's signal cannot reach Alice if he is at a distance greater
than $v_{g}T$ from Alice. The paradox is thus well-posed, the contradiction been now with FGV, not with relativity.

We now come to part 2.). Schr\"{o}dinger's principle implies that any measurement on Bob's spin or Alice's spin is to be regarded as
a measurement \textbf{on the entire two-spin system}. When Alice finds that her spin points ``up'', she is measuring the observable
$A= \mathbf{1} \otimes \sigma^{3}_{0}$ on the state $\omega_{B}$, i.e., she obtains the value $\omega_{B}(A) = 1$. Therefore, by assumption 2),
a collapse of the state $\Psi_{B,-}$ to the state $|-)_{1} \otimes |+)_{2}$ took place, whereby she actually learns that the 
\textbf{two-spin system} is in the state $|-)_{1} \otimes |+)_{2}$. 

Consequently, she does learn \textbf{at the same time} that \textbf{Bob will be necessarily measuring the value -1/2}. This result strongly
contrasts with EPR's version of the experiment, which may be described as follows: at the moment of her measurement, Alice is able to
\textbf{deduce} from her knowledge of the initial state ~\eqref{(1.2)} that Bob will be measuring the value $-1/2$, thereby obtaining
a \textbf{new} information, which, as we have previously shown, violates FGV if Bob is sufficiently far away.   

\end{proof}
\end{proposition}

\section{The nature of the measurement process and a simple generalization to free photon fields defined on Fock space}

As remarked in the introduction, our contribution in this paper was only to  show that the solution in Schr\"{o}dinger's paper
\cite{Schr} of the EPR-locality paradox, within the Copenhagen interpretation, may be made completely precise. Griffiths'
solution (\cite{Gri1}, \cite{Gri2}) was obtained within his probabilistic approach, but Griffiths does not
make explicit use of the collapse of the wave-packet, as he himself remarks in \cite{Gri2}.

At the same time, we have seen that certain important details of measurement theory are crucial: the condition $T < \infty$ on the time of 
measurement in the proposition, but also implicit is the existence of an \textbf{interaction} between the system and the measurement
apparatus, in \cite{Wre2} taken as an inhomogeneous external magnetic field in the case of a measurement of spin. 

A simple generalization covers free relativistic fields, e.g. photon fields, defined on Fock space 
\begin{equation}
\label{(3.1)}
{\cal F} = \oplus {\cal F}_{n}
\end{equation}

where ${\cal F}_{n}$ denotes the $n-$ particle subspace of Fock space ${\cal F}$. correspondingly, a state $\omega$ on ${\cal F}$ may be written
\begin{equation}
\label{(3.2)}
\omega = \{\omega_{n}, n=0,1, \cdots \}
\end{equation}

Above, $\omega_{n}$ denotes the restriction of the state $\omega$ to the $n-$ particle subspace ${\cal F}_{n}$, and $\omega_{0}$ is the
Fock vacuum (zero-particle subspace). A suitable generalization of the state $\omega_{B}$  describes two-photon entangled states
(\cite{Sa}, p.222) associated to the photon-polarization measurement. It also follows from the latter analysis that a choice of
``measurement apparatus'' similar to the inhomogeneous external magnetic field of Stern-Gerlach type chosen in \cite{Wre2}, but involving
photons with polarization-dependent momenta, is applicable.

\section{Local measurements and the necessity of interacting quantum fields}
\subsection{Measurements require interacting quantum fields}
 
If one wishes to describe a \textbf{local} measurement, the inclusion of (algebraic) relativistic quantum field theory (AQFT) becomes necessary.

The subject has been studied extensively by Haag \cite{Ha} and Fewster and Verch \cite{Verch}, along lines similar to the present paper, but,
as far as we know, no rigorous or exact results concerning the dynamics exist, for instance, in
the case of a local current, with, as natural interaction, the interaction with a local photon field, as in relativistic 
quantum electrodynamics, which is of the form
\begin{equation}
\label{(4.1)}
H_{I} = J^{\mu}(f)A_{\mu} 
\end{equation}
Above, the covariant notation is used, viz. the sum over the indices $0, \cdots, 3$, with zero denoting the time index, and $f$ is a smooth
function of compact support on space-time. In full space, for fixed time, the corresponding interacting density would be
\begin{equation}
\label{(4.2)}
H_{I}(\vec{x}) \equiv \int d\vec{x} J^{\mu}(\vec{x}) A_{\mu}(\vec{x})
\end{equation}
which characterizes (relativistic) quantum electrodynamics (qed). Thus: in the important example of the measurement of a \textbf{local} current,
the natural ``measuring apparatus'' is the photon field, and the theory is necessarily that of \textbf{interacting} quantum fields! We shall
assume that some form of algebraic relativistic quantum field theory (AQFT) is applicable to qed.

\subsection{Existence and general properties of entangled states: the ``analog'' of Schr\"{o}dinger's principle}

A seminal theorem of Clifton and Halvorson \cite{CH} asserts that the set of entangled states (a suitable generalization of the concept introduced
in section 2 in the proper topology for the states, see \cite{CH}) in AQFT is generic, viz. open and dense in the set of all states. 

An explicit construction of entangled states in AQFT somewhat analogous to those in section 2 was done by Doplicher (\cite{Do}, pp. 415-419).
Therefore, in principle, Schr\"{o}dinger's principle could apply to them. It is certainly not the form \eqref{(3.1)}, \eqref{(3.2)}, because,
by a corollary of Haag's theorem, an interaction density of the form \eqref{(4.2)} would have to annihilate the Fock vacuum (zero-particle state),
which does not happen due to the photon creation term in \eqref{(4.2)}, which polarizes the vacuum (for the complete argument, which has seldom
appeared explicitly, see Wightman's lectures in \cite{Wight}). Thus, we have to deal with a non-Fock representation. The standard way out is to
consider a cutoff Hamiltonian $H(\Lambda, \Delta)$, where $\Lambda$ denotes a space cutoff, and $\Delta$ an ultraviolet cutoff on the photon
field. This is the subject of the last section.

On the other hand, Doplicher's construction allows a partial answer to what would be the analogue of Schr\"{o}dinger's principle in the case of
measurement of local currents in qed: the analogues of ``two-particle states'' in Fock space correspond to local creation of electron-positron
pairs from the (interacting) vacuum: this is forced by global \textbf{charge neutrality}, viz. the necessity of remaining in the 
\textbf{vacuum sector} of the theory. The analogs of the entangled "two-particle states" are now entangled "electron-positron states" of the general form
\begin{equation}
\label{(4.3)}
\alpha a_{d}(f)^{*} b_{d}(g)^{*} \Omega + \beta a_{d}(g)^{*} b_{d}(f)^{*} \Omega
\end{equation}
where $\alpha$ and $\beta$ range in $[0,1]$ and $\alpha + \beta = 1$, and $a_{d}(f),b_{d}(g)$ denote electron and positron annihilation operator-valued distributions
acting on smooth functions $f,g$ with bounded support on space-time.  $\Omega$ denotes the vacuum of the interacting theory. Unfortunately, equation \eqref{(4.3)}
has a clear meaning only for free fields defined on Fock space. There, the index "d" meaning "dressed" may be ignored. Otherwise, the meaning of
\eqref{(4.3)} is symbolic, see the following section.

\subsection{Why perturbative AQFT in the form of formal series may fail for qed}

We now come back to the Hamiltonian $H(\Lambda, \Delta)$ of the previous section and inquire into the remaining possibility, that eventually 
perturbative AQFT \cite{BFR} might apply to qed. 

For nonrelativistic qed, the fact that the answer is \textbf{negative} may, perhaps surprisingly, be formulated
as a theorem.

\begin{theorem}
\label{Theorem 4.3}

Perturbative nonrelativistic qed in the guise of a formal series does not describe the properties of the Hamiltonian
$H(\Lambda,\Delta)$ correctly.

\begin{proof}
By the Lieb-Loss inequalities \cite{LL1}, if $E(\Lambda,\Delta)$ denotes the lowest (nonnegative) eigenvalue of $H(\Lambda, \Delta)$ (which is a non-negative
operator, see \cite{LL1}),
\begin{equation}
\label{(4.1)}
E(\Lambda, \Delta) \le c_{\Lambda} |\Delta|^{\frac{12}{7}}
\end{equation}

for sufficiently large $|\Delta|$, with $\Lambda$ fixed, where $c_{\Lambda}$ is a strictly positive constant (due to the lower bounds in \cite{LL1}).
On the other hand, perturbation theory in the form of formal series yields under the same conditions (see also \cite{Lieb} and references given there):
\begin{equation}
\label{(4.2)}
E(\Lambda, \Delta) = c_{\Lambda}|\Delta|^{2}
\end{equation}
which contradicts \eqref{(4.1)}.
\end{proof}
\end{theorem}

What Lieb remarks in \cite{Lieb} is that perturbation theory cannot converge, because if it did, it would yield the wrong answer regarding
the high-photon-frequency behavior of the ground state energy. Formal series render this statement rigorous in the above sense. 

The results of Lieb and Loss in \cite{LL2} suggest the the above result remains true in full, relativistic qed. It means that the physical Hamiltonian $H$, 
i.e., the limit of $(H(\Lambda, \Delta)- E(\Lambda,\Delta))$, as $\Delta \to \infty$, followed by $\Lambda \to \infty$, in the proper sense 
(and within a non-Fock representation) does not yield a positive $H \ge 0$ if the prescription of perturbation theory is followed, as a consequence 
of the previous theorem, that is to say, \textbf{stability} is broken.

The above theorem means that the energy of the vacuum of the interacting theory includes the huge energy of all electron-positron pairs, which is
not accounted for by perturbation theory, rigorously defined as a formal series. Therefore \eqref{(4.3)} has to include in the operators $a(f), b(g)$
an unknown "dressing" by this cloud of photons!

The reason for this unpleasant fact is is summarized by Lieb's prophetic words 
in \cite{Lieb}: ``The physical picture that begs to be understood on some decent level
is that the electron is surrounded by a huge cloud of photons with an enormous energy. We are looking for small effects, called ``radiative
corrections', and those effects are like a flea on an elephant. Perturbation theory treats the elephant as a perturbation of the flea''.

\textbf{Acknowledgement} We should like to thank Elliott Lieb very heartily. He has been a mentor and a friend for several decades, and his
stimulating lectures \cite{Li} have been the point of departure of the present manuscript. Thanks also to Klaas Landsman for constructive
criticism of a previous version, and for reminding us of the important work of Haag on the subject, as well as of Fewster and Verch.
We also thank Pedro L. Ribeiro for remarks and corrections.

\end{document}